\documentclass[prl,twocolumn,showpacs]{revtex4}
\usepackage{graphicx,epsfig}
\usepackage{times,amsmath,amssymb,latexsym}
\usepackage{rotating}
\usepackage{stmaryrd}

\newcommand{\ket}[1]{\left|#1\right\rangle} 

\newtoks\nslashfraction\nslashfraction={.13}
\newcommand{\nslash}[1]{\,\setbox0\hbox{$#1$}\setbox0\hbox
to\the\nslashfraction\wd0{\hss\box0}/\box0}

\begin{document}

\title{Vortex nucleation as a case study of symmetry breaking in quantum systems}
\author{D. Dagnino$^{1}$, N. Barber\'an$^{1}$, M. Lewenstein$^{2,3}$,
and J. Dalibard$^{4}$}
\affiliation{(1) Dept. ECM, Facultat de F\'isica, Universitat de Barcelona,
E-08028 Barcelona, Spain\\
(2) ICFO - Institut de Ci\`encies Fot\`oniques, Parc Mediterani de la Tecnologia 08860 Barcelona, Spain\\
(3) ICREA-- Instituci\'o Catalana  de Recerca i Estudis Avan\c
cats, E-08010, Barcelona, Spain\\
(4) Laboratoire Kastler Brossel, CNRS, UPMC, Ecole Normale Sup\'erieure, 24 rue
Lhomond, 75005 Paris, France}

\begin{abstract}

Mean-field methods are a very powerful tool for investigating weakly interacting many-body systems in many branches of physics. In particular, they describe with excellent accuracy trapped Bose-Einstein condensates. A generic, but difficult question concerns the relation between the symmetry properties of the true many-body state and its mean-field approximation. Here, we address this question by considering, theoretically, vortex nucleation in a rotating Bose-Einstein condensate. A slow sweep of the rotation frequency changes the state of the system from being at rest to the one containing one vortex. Within the mean-field framework, the jump in symmetry occurs through a turbulent phase around a certain critical frequency. The exact many-body ground state at the critical frequency exhibits strong correlations and entanglement. We believe that this constitutes a paradigm example of symmetry breaking in - or change of the order parameter of - quantum many-body systems in the course of adiabatic evolution.
\end{abstract}

\date{\today}
\maketitle

In classical physics, examples of the usefulness of mean-field theory go back to the
``molecular field theory" of magnetism \cite{Weiss:1907}. 
In the classical world, symmetry changes (or breaking) are driven by thermal
fluctuations, and in the standard Landau-Ginsburg scenario are associated with
increase of classical correlations. In quantum physics, the paradigm example of
applicability of the mean field concerns a weakly interacting quantum Bose gas and
Bose-Einstein condensation \cite{pit}. The mean-field description of the gas assumes
that its ground state $\Psi$ is approximated by a product state $\Psi(\vec
r_1,\ldots,\vec r_N)=\psi(\vec r_1)\ldots\psi(\vec r_N)$, of essencially
uncorrelated particles forming a superfluid Bose-Einstein condensate with order parameter $\psi$.

Of particular interest for quantum gases are quantum phase
transitions and symmetry changes/breaking driven by quantum fluctuations.
A celebrated example is the superfluid to Mott-insulator transition of bosons in an
optical lattice \cite{Jaksch:1998}. Another example yet to be explored
experimentally is the case of a fast rotating gas, when the number of vortices
is similar to the number of particles, or equivalently angular momentum 
$L\sim N^2$ \cite{coo}. The ground state of the system is then a strongly correlated quantum 
liquid such as the Laughlin state,
analogous to those emerging in quantum Hall physics \cite{Yoshioka:2002}. Here, we
consider another situation, dealing with the case of a relatively slowly
rotating gas at the threshold of the nucleation of the first vortex. We show that
owing to the symmetries of the system, the many-body state at
nucleation is strongly correlated and characterize its properties.

The symmetry change/breaking that results from vortex nucleation has drawn a lot of
attention since the discovery of superfluids \cite{gri}. For quantum
gases, atoms are usually confined in an isotropic harmonic trap and experience
an extra quadratic potential rotating at angular frequency $\Omega$ 
(for a review see ref.7). From a theoretical point of view, the
vortex nucleation can be tackled by several techniques, ranging from a mean-field
approach based on the Gross-Pitaevskii equation \cite{Feder:2000,sin,kas}
to the investigation of the many-body energy eigenstates
\cite{but,Bertsch:1999,Smith:2000,Jackson:2000,dag,roma,parke}.
Within the mean field framework, standard textbooks \cite{pit} associate vortex
nucleation with thermodynamic instability. Above a critical rotation frequency 
$\Omega_c$, the odd solution $\psi$ of the Gross-Pitaevskii equation 
with a single vortex \cite{pit1,gro} has a lower energy than the even 
solution corresponding to the Bose-Einstein condensate at rest \cite{strin99}. Here, we go
beyond the mean-field approach and study the exact quantum dynamics of a
mesoscopic sample of atoms, in the presence of the stirring potential. Our main result
is that for a rotation frequency close to $\Omega_c$, the mean-field description is
invalid. The system enters a strongly correlated and entangled state, well described
by an effective two-mode model. We compare our results with those obtained from a
mean-field description and show that the latter  exhibits dynamical instability and
hysteresis. As we explicitly include here an anisotropic stirring potential, 
the present mechanism concerns  a discrete parity symmetry breaking. Therefore, it 
differs from the case of the vortex nucleation in axially symmetric traps: 
in the latter case, breaking of the continuous rotational symmetry involves 
a gapless Nambu-Goldstone mode \cite{Ued}, whereas here we deal with a gapped system. 

\paragraph{Model.}
We consider a mesoscopic sample of $N$  bosonic atoms 
of mass $M$ placed in an axially symmetric harmonic potential $V_0$, with frequency 
$\omega_\perp$ in the $xy$ plane and $\omega_z$ along
the $z$ axis. Here, $\hbar \omega_z$ is large compared with the interaction
energy so that the dynamics along $z$ is frozen and the gas is effectively
two-dimensional (2D) at sufficiently low temperature. The gas is set in rotation
using an anisotropic quadratic potential $V$ in the $xy$ plane, rotating at
angular frequency $\Omega$ around the $z$ axis. In the rotating frame, this stirring
potential reads $V(x,y)=2AM\omega_\perp^2(x^2-y^2)$, where the
coefficient $A$ ($\ll 1$) measures the strength of the anisotropy.

For $A\ll 1$ and $\Omega\sim \omega_\perp$, the single-particle energy levels in the
rotating frame are grouped in Landau levels, separated by $\hbar
(\omega_\perp+\Omega)$ (refs.7,22). We assume that $\hbar (\omega_\perp+\Omega)$ is large
compared with the interaction energy, so that the atomic dynamics is restricted to
the lowest Landau level (LLL). For $A=0$, a basis of the LLL single-particle states
is the set $\varphi_m(x,y) \propto (x+iy)^m e^{-(x^2+y^2)/2\lambda_\perp^2}$, where
$m\geq 0$ is an integer and $\lambda_\perp=\sqrt{\hbar/M\omega_\perp}$. Each
$\varphi_m$ is an eigenstate of the $z$-component of the single-particle angular
momentum (eigenvalue $m\hbar$) and of the single-particle Hamiltonian without
anisotropy (eigenvalue $\hbar [\omega_\perp+m (\omega_\perp-\Omega)]$). Within the
LLL, we model the atomic interactions by a 2D contact potential $U(\vec
r)=(\hbar^2 g/M)\,\delta(\vec r)$ where 
$g=\sqrt{8\pi}a/\lambda_z$ is dimensionless, $a$ is the 3D scattering length and $\lambda_z =
\sqrt{\hbar/M\omega_z}$. We
choose $\lambda_\perp$, $\hbar \omega_{\perp}$ and $\omega_{\perp}$ as units of
length, energy and frequency.

\paragraph{Energy spectrum.}
We first  recall some important properties of the $N$-particle system in
absence of anisotropy ($A=0$). In this case, the total angular momentum operator
$\hat L$ commutes with the Hamiltonian $\hat H$ so that one can look for the
eigenstates of $\hat H$ within subspaces ${\cal E}_L$ of fixed $L$.
The lowest-energy state in each ${\cal E}_L$ for $2\leq L\leq N$ is  
\cite{Bertsch:1999,Smith:2000,Jackson:2000}:
\begin{equation}
\Phi_L(\vec r_1,\ldots,\vec r_N) \propto \sum_{1\leq i_1 \ldots \leq i_L}
(u_{i_1}-u_c)\ldots (u_{i_L}-u_c)\; \Phi_0
\nonumber
 \label{eq:PhiL}
 \end{equation}
where $u_j=x_j+iy_j$, $u_c=\sum_j u_j/N$ and
 \begin{equation}
\Phi_0(\vec r_1,\ldots,\vec r_N)\propto e^{-\sum_j r_j^2/2}\ .
\nonumber
 \label{eq:Phi0}
 \end{equation}
The energy of the state $\Phi_L$ is $N+(1-\Omega) L+gN(2N-L-2)/(8\pi)$. At
$\Omega_1=1-gN/(8\pi)$, all $\Phi_L$ states for $L=0$ and $2\leq L\leq N$ are degenerate.
The angular momentum of the ground state $L_{\rm GS}(\Omega)$ shows sharp steps at critical
values $\Omega_i,i=1,2,..$ (ref.23). Below $\Omega_1$, the ground state is the
zero angular momentum state $\Phi_0$. At $\Omega_1$, $L_{\rm GS}$ jumps from $0$ to
$N$. Above $\Omega_1$ the ground-state  angular momentum has a plateau $L=N$ up to $\Omega_2$,
where a second jump takes place. From this value, a sequence
of jumps and plateaux emerges up to the last possible $L$ value, $L=N(N-1)$,
corresponding to the Laughlin state. In the following, we focus on the vicinity of
the first jump $\Omega \sim \Omega_1$, where the first vortex is nucleated.

\begin{figure}
\includegraphics[width=\columnwidth]{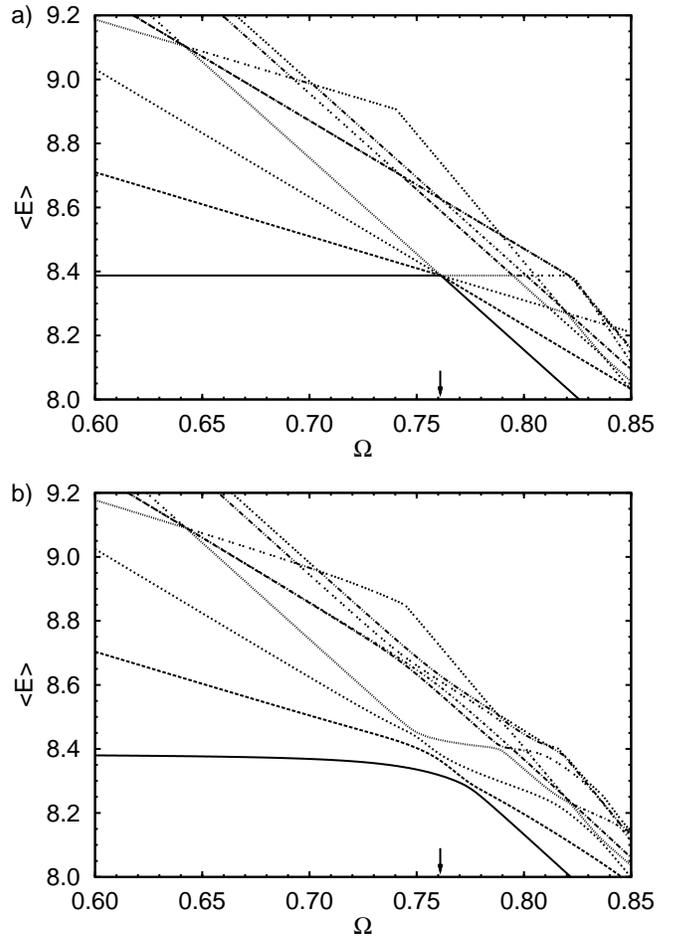}
\caption{\textbf{Energy spectrum as a function of $\Omega$. a}, Anisotropy parameter
$A=0$. \textbf{b}, $A=0.03$. In both cases, $N=6$ and $g=1$. For $A=0$, the ground state is multiply degenerate at the rotation
frequency $\Omega_1=1-gN/(8\pi)$, which corresponds to the nucleation frequency of
the first vortex. A non-zero anisotropy parameter lifts the degeneracy of the groun state.
Here, we plot only the first nine energy eigenvalues from the subspace formed with
even values of the total angular momentum, which are the only relevant ones for the
problem addressed in this article. The arrows mark the value of $\Omega_1$.}
 \label{fig:spectrum}
\end{figure}

We now turn to the case where the rotating anisotropy is present. The many-body
energy spectrum is calculated numerically by diagonalization of the Hamiltonian (see the
Methods section). We show it in Fig.1 for both  zero anisotropy and
for $A=0.03$, using $N=6$ for illustration. The interaction coupling $g=1$ so that $\Omega_1=0.761$. For
$A\neq 0$ the ground state does not show any degeneracy around $\Omega_1$, contrary to the
case $A=0$. In Fig.2, we compare $L_{\rm GS}(\Omega)$ for $A=0$
and $A=0.03$. For $A\neq 0$ $L_{\rm GS}$ evolves smoothly from $0$ to $N$ around $\Omega_1$.

\begin{figure}
\includegraphics[width=80mm]{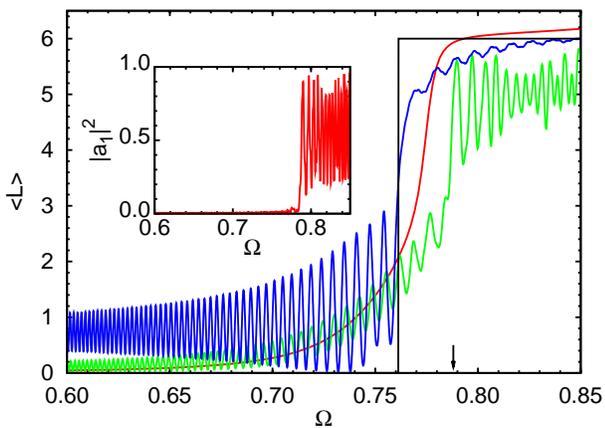}
\caption{\textbf{Variation of the angular momentum with rotation frequency $\Omega$}. The
black and red lines show, for an anisotropy $A=0$ and $A=0.03$ respectively, the
angular momentum of the ground state for a system of $N=6$ particles and an interaction
strength $g=1$. The green line is the average angular momentum predicted by the
mean-field treatment when $\Omega$ is linearly ramped from $0$ to $0.85$ with a
slope $\dot{\Omega}=10^{-4}$. The initial state at $\Omega=0$ is given by a slight
perturbation of the coefficients $a_0=1$, $a_1=a_2=0$. It presents a dynamical
instability of the zero-vortex mean-field solution for $\Omega=0.788$ (marked by an
arrow). Inset: The evolution of $|a_1|^2$, which explicitly shows
the instability. The blue curve of the main figure is the
backward evolution corresponding to an initial state at $\Omega=0.85$ close to the stationary mean-field solution 
$a_0=a_2=0$ and $a_1=1$. This solution ceases to exist for $\Omega<0.764$, causing the large oscillations in the evolution of the angular momentum.} \label{fig:Lz}
\end{figure}

\paragraph{Failure of the mean-field approach for $\Omega\sim\Omega_1$.}
We now explain why a mean-field description must fail at $\Omega \simeq \Omega_1$.
We notice  that the total Hamiltonian is parity invariant. Consequently, one can look
for an eigenbasis of the $N$-body Hilbert space composed of either even or odd
states. From the ground state of the Hamiltonian, we can extract the single-particle density
matrix (SPDM) $n^{(1)}(\vec r,\vec r')$ (see the Methods section), which is also parity
invariant. Hence, the single-particle orbitals $\psi_k$, which are eigenstates of
$n^{(1)}$ with eigenvalues $n_k$ ($\sum_k n_k=N$), can also be chosen with even or
odd parity. Suppose that we vary $\Omega$ from an initial value $\Omega_i$
($\Omega_i <\Omega_1$) to a final value $\Omega_f$ ($\Omega_1<\Omega_f<\Omega_2$),
choosing $\Omega_{i,f}$ in a region where the mean-field description is valid, that is,
when the largest eigenvalue $n_1$ is close to $N$. For $\Omega_i<\Omega_1$ the most
(second most) populated state $\psi_1$ ($\psi_2$) has no (has a) vortex in its
central region and is even (odd).  Choosing $\Omega_i=0.7$, we plot the phase
profiles of $\psi_{1,2}$ in the first row of Fig.3 for $N=6$
atoms, $g=1$ and $A=0.03$. On the other hand, at $\Omega_f$  the ground state has a single
well-centered vortex and $\psi_1$ and $\psi_2$ are odd and even, respectively
(see last row of Fig.3 for $\Omega=0.8$). Hence, the parity
of $\psi_1$ must change at some intermediate $\Omega_c$, which is close (for small
$A$) to the vortex nucleation frequency $\Omega_1$ in absence of anisotropy. By
continuity, the two most populated eigenstates $\psi_1$ and $\psi_2$ of $n^{(1)}$
must have equal populations, heralding a failure of the mean-field at $\Omega_c$.

We show in Fig.4 the variation of $n_1/N$ and $n_2/N$ as a function
of $\Omega$, for $N=12$, $g=0.5$ and $A=0.03$. These two populations are equal for
$\Omega_c=0.775$. We see that $n_1+n_2\simeq N$ over the whole range of frequencies
of this figure, indicating that most of the population of the SPDM is concentrated
in the first two modes $\psi_1$ and $\psi_2$. We checked up to $N=20$ that this 
concentration increases with $N$. Another relevant fact is that only
the first three LLL single-particle states ($m=0,1,2$) have a significant weight in
the expansion of $\psi_1$ and $\psi_2$. More specifically, below $\Omega_c$ $\psi_1$ is 
approximately a coherent superposition of $\varphi_0$ and $\varphi_2$, 
corresponding to two off-centered vortices (even parity), whereas $\psi_2$ is very close to a well-centered 
single-vortex state $\varphi_1$ (odd parity). Above $\Omega_c$, $\psi_1$ and $\psi_2$ abruptly 
exchange their form (see Fig.3). 

\begin{figure}
\includegraphics[width=\columnwidth]{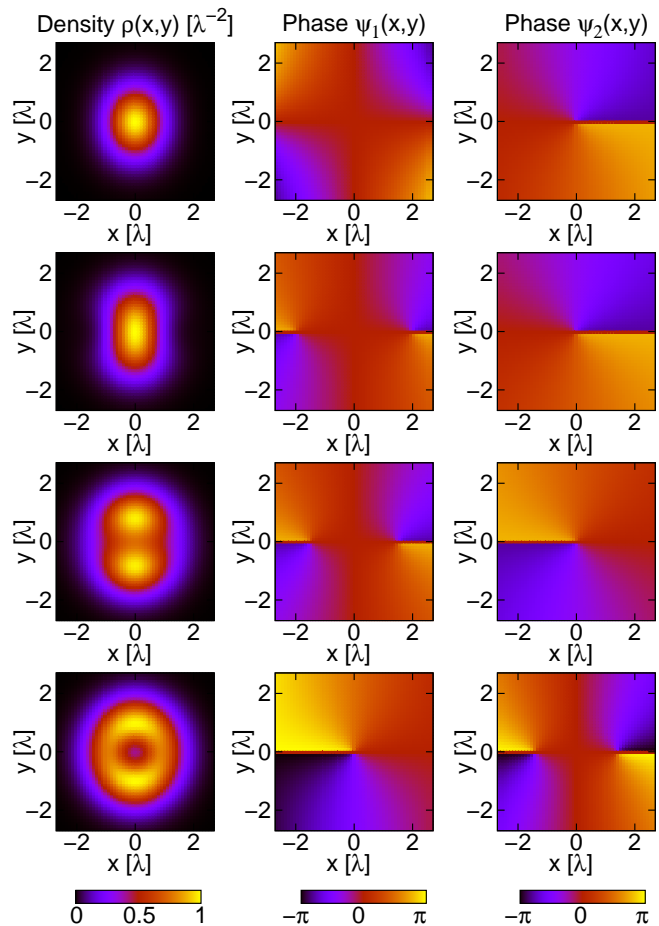}
\caption{ \textbf{Density of the ground state and phase maps of $\psi_1$ and $\psi_2$}. Four different values of $\Omega$ for $N = 6$, $A = 0.03$ and $g = 1$ are considered. First row: $\Omega=0.7$, $n_1 = 5.85$, $n_2 =0.12$.
Second row: $\Omega=0.760$, $n_1=5.01$, $n_2=0.60$. Third row:
$\Omega=\Omega_c=0.776$ $n_1=n_2=2.88$. Fourth row: $\Omega=0.8$ with $n_1 = 4.24$,
$n_2 = 1.07$. The first column is the contour plot of the total density, and the
second and third columns show the local phase maps of $\psi_1$ and $\psi_2$
respectively. Vortices are localized at the singularities of the phase maps,
surrounded by diffuse change of the phase. This figure shows that the nucleation of
the first centered vortex in a rotating condensate by a slow frequency sweep does
not occur through a smooth entrance of the vortex. The system passes through a
correlated, non-mean-field state where two single-particle states have equal weight.
At this point, $\psi_1$ changes from being a coherent superposition of $\varphi_0$ 
and $\varphi_2$ (two off-centered vortices) to the single $\varphi_1$ state, which 
corresponds to a well-centered single vortex. Simultaneously, $\psi_2$ experiences 
the inverse change.}
 \label{fig:two_points}
\end{figure}

\begin{figure}
\includegraphics[width=\columnwidth]{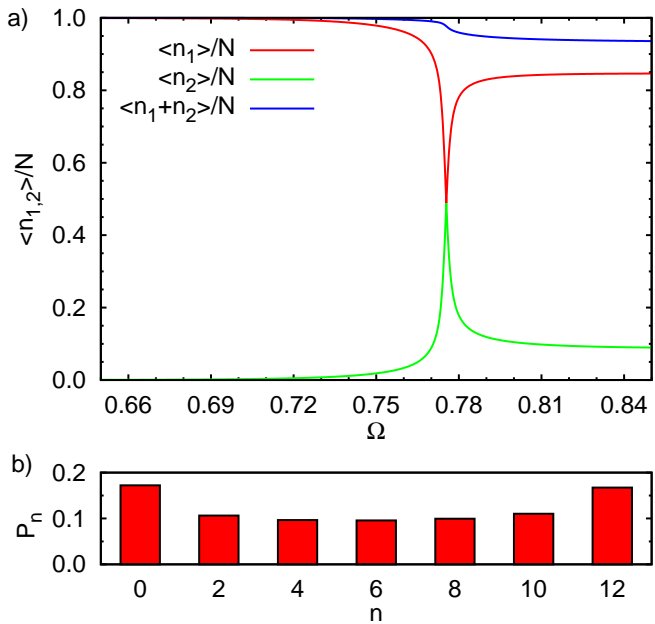}
\caption{\textbf{Structure of the ground state. a}, Variation of the relative populations $n_1/N$ and $n_2/N$ of the two
most occupied states $\psi_1$ and $\psi_2$ of the SPDM. When $\Omega$ is
sufficiently different from $\Omega_c$, $n_1\simeq N$, the system is well described
by a single mode and the mean-field description is valid. Conversely, for
$\Omega\simeq \Omega_c$, the two populations are comparable, corresponding to the case where  a two-mode approximation is valid even in the entangled region. \textbf{b}, Analysis of
the state of the system at the critical point where $n_1=n_2$, in terms of the square of the
scalar products $P_n=\mid \langle n:\psi_1\; ;\; N-n:\psi_2|\Psi_0\rangle \mid ^2$.
We obtain $\mid\langle E\mid \Psi_0\rangle\mid = 0.92$ (see equation (1)). Both panels are
plotted for $N=12$, $g=0.5$ and $A=0.03$.}
 \label{fig:n1n2}
\end{figure}

The failure of the mean-field description around $\Omega_c$ may occur in two ways. A first 
possibility is that for $\Omega=\Omega_c$, the many-body ground level itself has  a 
two-fold degeneracy with two eigenstates of opposite parity. This scenario corresponds 
to a first-order transition. It occurs when $N$ is
odd, because the graund state evolves from $\sim \psi_1^{\otimes N}$ with $\psi_1$ even to $\sim
\psi_1^{\otimes N}$ with $\psi_1$ odd.
The second possibility is that the many-body graund state $\ket{\Psi_0}$ remains non-degenerate, as this is the case in Fig.1b. In this case,
$\ket{\Psi_0}$ is even over the whole range $[\Omega_i,\Omega_f]$. This occurs for
even $N$ and will be of interest for the rest of the article. 

\paragraph{Quantum correlations for $\Omega\sim \Omega_c$.}
We have carried out a detailed  study of the ground state $\ket{\Psi_0}$ 
around the critical frequency $\Omega_c$, where the two largest eigenvalues of the
SPDM are equal ($n_1=n_2$). At criticality, the system is very well described by a
two-mode approximation implied by Fig.4a. The two largest
eigenvalues of the SPDM are much larger than all of the others, so that $n_1=n_2\simeq
N/2$. For example, for $N=12$, $g=0.5$ and $A=0.03$, we obtain $n_1=n_2=0.49\,N$ at
$\Omega_c=0.776$. The ground state is strongly correlated and is well described (for
even $N$) by
\begin{equation}
\ket{E}=[\ket{N,0}+\ket{N-2,2}+...+\ket{0,N}]/\sqrt{N/2+1}, \label{our}
\end{equation}
where $\ket{n,m}$ is the state with $n$ (respectively $m$) atoms in $\psi_1$ (respectively
$\psi_2$). Amazingly, the form of the ground state at $\Omega_c$ is practically independent of $A$, as long as $A\ll 1$. 
For a quantitative comparison of the exact ground state with the state (1),
we show in Fig.4b the squared scalar products $\langle n, N-n|\Psi_0\rangle$
in the case $N=12$. They are all zero for odd values of $n$ (as expected from the
parity of $\ket{\Psi_0}$) and approximately constant for even values of $n$. We
compared also our ground state at $\Omega_c$ with other celebrated correlated
states, such as ``Schr\"odinger cat" states $(\ket{N,0}+\ket{0,N})/\sqrt 2$, or
``twin" states $\ket{N/2,N/2}$, and found much smaller overlaps. 
Although there are various ways of defining entanglement for identical
particles (for a review see \cite{eckert}), according to Zanardi's concept of  mode
entanglement (ref.24), the state (1) is maximally entangled. This is
clearly seen by tracing the state (1) over one of the two modes and observing that
the von Neumann entropy of the reduced density matrix reaches the
maximal value $S\sim \log(N)$. 

At this point we mention related work on rotating ring lattices and Josephson 
junctions \cite{nun}. There, strongly correlated states are predicted at critical 
rotation, but the mechanism of their generation, as well as their nature 
are fundamentally different. The starting situation of these discretized models 
is that there are two degenerated  single-particle states. Interactions lift the 
degeneracy in the many-body system and favor the ``cat" states. In our case, the ground state 
for $A=0$ is macroscopically degenerated in the presence of interactions. 
The degeneracy is lifted here by the anisotropy, leading to another kind of strongly 
correlated ground state.

\paragraph{Vortex nucleation with adiabatic passage.}

We now study the real-time dynamics of the system using the time-dependent
Schr\"{o}dinger equation. A quasi-adiabatic evolution that brings the system from the zero-vortex to the
one-vortex state, is realized  by sweeping
$\Omega$: $\Omega(t)= \Omega_{\rm i} + \gamma t$ from the initial
frequency $\Omega_i$ chosen well below $\Omega_c$ (typically $\Omega_i=0.65$) to
the final frequency $\Omega_{\rm f}$, well above $\Omega_c$ (typically
$\Omega_f=0.85$). This evolution produces as an intermediate
step the strongly correlated state (1). The key parameter for the success of
this quasi-adiabatic evolution is the energy gap $\Delta$ between the ground state
and the first excited state of the system.

We have carried out a study of this gap for various $N$, keeping the product $Ng$
constant so that $\Omega_1$ also remains constant. We found that for small $A$ values
(below 0.1), the gap is roughly constant over the range $10\leq N \leq 20$, and
equal to $\simeq 0.5\;A$. Knowing the gap, we estimate the largest
possible $\gamma$ compatible with adiabatic evolution following ref. 27 and find 
$\gamma_{\rm max} =\xi\; \Delta^2/N$, where $\xi\ll 1$ (see the Methods section). This criterion agrees well with our
results. Defining as successful an adiabatic evolution that leads to an overlap
larger than 0.98 between the final state and the  ground state at $\Omega_f$, we find
$\xi\simeq 0.1$ for $10\le N\le 20$. Such a quasi-adiabatic evolution anables us  
to attain the correlated state (1) with comparable overlap.
For practical implementations, the atoms can be confined in a relatively tight trap
at the nodes of an optical lattice with $\omega_\perp/2\pi$ in the 10 kHz range. For
an anisotropy $A=0.1$ and $N=10$ atoms, the sweep time has to be of the order of one
second to ensure adiabaticity.

A natural question is the generalisation of the present scheme to large $N$. Assuming 
that the gap protecting the ground state remains constant, the mechanism will in principle survive. 
However, we have neglected here any parity-breaking perturbation in the Hamiltonian. 
Such a term would couple the subspaces corresponding to even and odd $L$ values. As shown 
in ref. 17, the lowest energies of these two subspaces are exponentially close when 
$N$ increases, wich affects the robustness of the ground state. This
coupling thus constitutes an important decoherence mechanism for large $N$,
whereas our scheme remains valid for $N$ not exceeding a few tens.

\paragraph{Mean field approach.}

As our results point out that strongly correlated states may be reached in
the course of the time evolution, it is interesting to see what the predictions of
usual mean-field theory are. To this aim, we expand the
condensate wavefunction $f(\vec r,t)$ into the relevant single
particle LLL orbitals $\varphi_m(\vec r)$ with angular momentum $m=0,1,2$,
$f(\vec r,t)=\sum_{m=0}^2a_m(t)\varphi_m(\vec r)$. Using the dynamical
variational principle \cite{perez}, we derive Lagrange equations for the complex
amplitudes $a_m(t)$ (see the Methods section), and look for the stationary solutions of the form 
$a_m(t)=\exp(-i\mu t)a_m(0)$ and their
stability. Finally, we evolve the mean-field equations and
compare the results with the full quantum treatment.

We choose $gN=6$ and $A=0.03$. Among the several possible stationary
solutions, two of them are relevant. The
first one $f_a$  corresponds to the ``no vortex" situation with a small admixture of
``two-vortex" orbital, that is, $|a_0|\simeq 1,|a_2|\ll 1$ and $a_1=0$. This solution
is the ground state for $\Omega<\tilde{\Omega}=0.773$. The second relevant
solution $f_b$ contains a non-zero contribution of the one-vortex state ($a_1\neq 0$) and
it is the ground state for $\Omega>\tilde{\Omega}$. Thus, $\tilde{\Omega}$ marks the critical
value of $\Omega$ for the thermodynamical stability of a centred vortex,
and the ``first-order transition" within the mean-field approach.

In the frequency range $0.764<\Omega<0.788$, both solutions exist and
are stable, leading to a bistable and hysteresis behaviour. 
For $\Omega>0.788$, $f_a$ becomes dynamically
unstable ($|a_1|$ grows exponentially in time, starting from noise, see inset in Fig. 2).
For $\Omega<0.764$, $f_b$ does not exist. The
numerical study confirms this hysteresis
behaviour, as shown in Fig. 2. The green line shows the angular
momentum  when $\Omega$ is ramped linearly in time from
$\Omega_i=0$ to $\Omega_f=0.85$, with the rate $\dot \Omega=10^{-4}$. A turbulent
behaviour occurs once $\Omega(t)$ reaches the edge of the 
stability domain of $f_a$. The blue line shows the reverse evolution in which $\Omega$ varies
from $\Omega_f$ to $\Omega_i$ at the same rate. Evidently, 
the adiabatic character of the dynamics cannot be maintained,  in contrast to
the result of the exact many-body treatment.

\paragraph{Summary}
We conjecture that the scenario presented above is generic for the following situations:
(1) it concerns quantum mechanical systems in which the ground state undergoes
symmetry change/breaking as some parameter of the system
$\lambda$ crosses a critical value $\lambda_c$; (2) far from $\lambda_c$, the systems are well described by the
mean-field theory with order parameters reflecting the change of
symmetry; (3) in the dynamical mean-field description, the system exhibits dynamical
instability and breakdown of adiabaticity.

In such situations we expect the appearance of strongly correlated states. The SPDM
shows typically a few relevant single-particle modes
that are involved in the symmetry change. They can be guessed
by analyzing the results of the  dynamical mean-field approach. For
instance, if this approach exhibits standard signatures of bistability, we can
expect two relevant modes as in the case study presented here. Similar insight can
be gained from analysis of small Gaussian fluctuations around the mean-field
solutions, that is, Bogoliubov-de Gennes equations \cite{garay}. Reduction of the full
theory to the quantum modes provides then a very good approximation.
Alternatively, it can be viewed as re-quantization of the mean-field theory reduced
to the relevant single-particle orbitals \cite{parke}. The strongly correlated
states appearing in such a situation exhibit strong entanglement and this 
property can be detected in experiments with moderate $N$.

\section*{Methods}

\paragraph{Diagonalization of the Hamiltonian.}

In the frame rotating at angular frequency $\Omega$, the Hamiltonian of the system
is $H=H_0+U$, where $H_0$ is the sum of one-body Hamiltonians $H_0=\sum_{j=1}^N
H_{0,j}$ and $U$ is the two-body interaction potential, characterized by the
3D scattering length $a$. Each one-body Hamiltonian is the sum of
kinetic, potential and rotation energy:
\begin{eqnarray}
H_{0,j} & = & \frac{p_j^2+p_{zj}^2}{2M}+\frac{M}{2}(\omega_{\perp}^2r_j^2+
\omega_{\parallel}^2 z_j^2) \nonumber
\\
& - & \Omega L_{z,j} +V_{j}
\nonumber
\end{eqnarray}
where $V_{j}=2AM\omega_{\perp}^2(x_j^2 - y_j^2)$ is the anisotropic potential that
sets the gas in rotation. We assume that the interaction energy is much smaller than
$\hbar \omega_z$ so that the $z$ motion is frozen and the atoms occupy only the
ground state $\exp(-z^2/(2\lambda_z^2))$ of this degree of freedom. The gas is
supposed to be rotating sufficiently fast to have $\omega_\perp-\Omega \ll
\omega_\perp+\Omega$, which guarantees that the various Landau levels are well separated
from each other. The interaction energy is also assumed to be small compared to
$\hbar (\omega_{\perp}+\Omega)$ so that the low temperature dynamics is restricted to the
LLL.

In the absence of anisotropic potential $A=0$, the eigenstates of the
one-body Hamiltonian in the LLL are the functions $\varphi_m(x,y) \propto (x+iy)^m
e^{-(x^2+y^2)/2\lambda_\perp^2}$, $m=0,1,2,\ldots$. We introduce the creation
$a_m^\dagger$ and annihilation $a_m$ operators of an atom in state $\varphi_m$, and
we write $H$ in the second quantization
\begin{equation}
\hat{H}=\hbar \,\, \omega_\perp \hat{N}+\hbar \,\,(\omega_{\perp}-\Omega) \hat{L}+\hat{V} + \hat{U}\,\,\,,
\nonumber
\end{equation}
where $\hat{N}=\sum a^\dagger_m a_m$ and
$\hat{L}=\sum m a^\dagger_m a_m$ are the particle number operator and the total
$z$-component angular momentum operator, respectively. The expression of the
rotating potential in the second quantization is
\begin{eqnarray}
\hat{V} & = &\frac{A}{2} \lambda_\perp^2\sum_m \Big(\sqrt{m(m-1)}\;a^{\dag}_m\,a_{m-2}
\nonumber
\\
& + &\sqrt{(m+1)(m+2)}\;a^{\dag}_m\,a_{m+2}\Big)\ .
\nonumber
\end{eqnarray}
Finally the contact interaction potential reads
 \begin{equation}
\hat{U}=\frac{1}{2}\sum_{m_1m_2m_3m_4}
U_{1234}\,\,\,a^\dag_{m_1}a^\dag_{m_2}a_{m_4}\,a_{m_3}\ ,
\nonumber
 \end{equation}
where the matrix elements are given by
\begin{eqnarray}
U_{1234}& = &\langle m_1\,m_2 \mid U \mid m_3\,m_4 \rangle \nonumber
\\
& = &\frac{g}{\lambda_\perp^2 \pi}\,\,\frac{\delta_{m_1+m_2,
m_3+m_4}}{\sqrt{m_1!m_2!m_3!m_4!}}\,\, \frac{(m_1+m_2)!}{2^{m_1+m_2+1}}\: .
\nonumber
\end{eqnarray}

In the absence of anisotropy ($A=0$), $\hat{H}$ and $\hat{L}$ commute and share a common
basis. The first step in the diagonalization of the Hamiltonian is to determine a
basis $|\Lambda_p\rangle$ ($p=1,\ldots,n_L$) for each subspace of given total angular
momentum $L$. The dimension $n_L$ of each subspace corresponds to all of the possible
configurations of $N$ particles with angular momentum $m_j$ that fulfil the
condition $L=\sum_{j=1}^N m_j$. The matrix of the Hamiltonian in the LLL basis
then consists of blocks of size $n_L\times n_L$, which we diagonalize using standard
codes.

When $A\neq 0$, the anisotropic potential connects the various subspaces of given
$L$. We then choose a maximum angular momentum $L_{\rm max}$ and write the matrix
giving the restriction of the Hamiltonian to the subspace of states with $L\leq
L_{\rm max}$. This $Q\times Q$ matrix, with $Q=\sum_{L=0}^{L_{max}} n_L$, is again
diagonalized using standard codes. In practice the value of $L_{\rm max}$ is chosen
to ensure a good convergence for the energies and the eigenstates of the
Hamiltonian. The results given here have been obtained with $L_{\rm
max}=N+2$.

Note that the anisotropic rotating contribution $V$ can in principle be included
within the framework of the LLL approximation in two ways. The first approach has
just been described above and consists of keeping the same Landau levels as for
$A=0$ and then diagonalizing $\hat H$ within the LLL.
The second approach consists of calculating exactly the single-particle eigenstates 
in presence of the anisotropy $V$, and defining a new LLL
accordingly \cite{Fetter:2007}. The Hamiltonian is then diagonalized within this
`anisotropic' LLL. We have checked that both methods lead to very similar results
for $\Omega \sim \Omega_1$. The results presented here have been obtained
with the first approach.

\paragraph{Single particle density matrix }
The SPDM can be regarded as an integral operator with the kernel:
\begin{equation}
n^{(1)}(\vec{r},\vec{r'}) = \langle \Psi_0\mid \hat{\Psi}^{\dag}(\vec{r})\,
\hat{\Psi}(\vec{r'})|\Psi_0 \rangle,
\nonumber
\end{equation}
with $\hat{\Psi}(\vec r)$ and $\hat{\Psi}^\dag$ being the annihilation and creation
field operators of an atom in $\vec r$. The single-particle orbitals are the
eigenstates of the SPDM:
\begin{equation}
\int d\vec{r'} n^{(1)}(\vec{r},\vec{r'}) \psi^{*}_k(\vec{r'})= n_k
\psi_k(\vec{r}).
\nonumber
\end{equation}
If there exist a single relevant eigenvalue such that $n_1 \gg \sum_{k\geq 2 }n_k$,
then $ \sqrt{n_1} \psi_1(\vec{r})$ has the role of the order parameter of the
system. In particular, the map of the local phase of this function gives precise
information on the location of vortices \cite{dag}.

\paragraph{Adiabatic approximation}
The diagonalization of the many-body Hamiltonian provides the eigenstates
$|\Psi_j(\Omega)\rangle$ and the eigenenergies $E_j(\Omega)$. In particular, the
ground state $|\Psi_0(\Omega)\rangle$ is separated from the first excited state
$|\Psi_1(\Omega)\rangle$ by an energy gap $\hbar \omega_{10}(\Omega)$, which is minimal
at the avoided crossing close to $\Omega_1$. We consider here a process where
$\Omega$ is scanned linearly from $\Omega_i<\Omega_1$ to $\Omega_f>\Omega_1$ and we
want to find a criterion on $\dot \Omega$ ensuring that the system follows
adiabatically the  ground state, with a negligible transition rate to the other states.

The probability for a non-adiabatic transition $\Psi_0 \rightarrow \Psi_j$ is given
by \cite{Messiah}:
 \[
p_{0\to j}\leq \mbox{max}\left(\frac{\alpha_{j0}}{\omega_{j0}} \right)^2
 \]
where $\alpha_{j0}=\langle\Psi_j | (d|\Psi_0\rangle/dt)$. We have
 \[
\frac{d|\Psi_0\rangle}{dt}=\dot \Omega\;\frac{d|\Psi_0\rangle}{d\Omega} .
 \]
From the eigenvalue equation $H|\Psi_0\rangle = E_0|\Psi_0\rangle$, we obtain after
a derivative with respect to $\Omega$:
 \[
-L_z \; |\Psi_0(\Omega)\rangle+ H(\Omega)\frac{d|\Psi_0\rangle}{d\Omega}=
 \frac{dE_0}{d\Omega}\;|\Psi_0(\Omega)\rangle\ +\ E_0\frac{d|\Psi_0\rangle}{d\Omega}
 \ .
 \]
We now take the scalar product with $\langle \Psi_j|$ ($j\neq 0$) and we get:
 \[
\langle \Psi_j|L_z|\Psi_0\rangle=(E_j-E_0) \langle \Psi_j|
\frac{d|\Psi_0\rangle}{d\Omega}\ .
 \]
We choose $|\Psi_j\rangle$ equal to the first excited state of the system
$|\Psi_1\rangle$. The matrix element $\langle \Psi_1|L_z|\Psi_0\rangle$ is at most
of order $N\hbar$ in the vicinity of the avoided crossing. Therefore:
 \[
\alpha_{10}=\langle\Psi_1 | \frac{d|\psi_0\rangle}{dt} \leq \dot
\Omega\;\frac{N\hbar}{\hbar \omega_{10}}\ ,
 \]
hence the condition for $p_{0\to 1}\ll 1$:
 \[
\dot \Omega \; \frac{N}{\omega_{10}^2}\ll 1\ .
 \]

\paragraph{Mean-field approach.}

The mean-field approach consists of assuming that all atoms are in
the same state $f(\vec r,t)=\sum_{m=0}^2a_m(t)\varphi_m(\vec r)$ with $\sum
|a_m|^2=1$. The average angular momentum per particle is $L=|a_1|^2+2|a_2|^2$ and
the average energy per particle $E(\psi)=\frac{1}{N}\langle f^{\otimes
N}|H|f^{\otimes N}\rangle$ reads (up to an additive constant):
\begin{eqnarray}
E(\psi)&=&(1-\Omega) (|a_1|^2+2|a_2|^2)+\sqrt{2}A(a_0a_2^*+a_0^*a_2) \nonumber \\
&+& \frac{Ng}{4\pi} \Big[ |a_0|^4 + \frac{1}{2}|a_1|^4 +\frac{3}{8}|a_2|^4
 \nonumber \\
&&+2|a_0|^2|a_1|^2+|a_0|^2|a_2|^2+\frac{3}{2}|a_1|^2|a_2|^2 \nonumber
\\
&&+\frac{1}{\sqrt{2}}
 (a_0a_2(a_1^*)^2+ a_0^*a_2^*a_1^2)
\Big]\ .
\nonumber
\end{eqnarray}
The Lagrange equations associated with this energy are $i\dot a_j =\partial
E/\partial a_j^*$ (ref. 28), which gives for example:
 \[
i \dot{a}_0=\sqrt{2} A a_2 + \frac{Ng}{2\pi}\Big[ a_0\Big(|a_0|^2+
|a_1|^2+\frac{1}{2}|a_2|^2\Big)+\frac{1}{2\sqrt{2}}a_1^2a_2^*\Big]
 \]
and two similar equations for $\dot a_1$ and $\dot a_2$. Note that in this mean-field
approach, $N$ and $g$ have a role only through the product $Ng$. In particular, the
fact that $N$ is even or odd is of no relevance here.

The stationary solutions are obtained by inserting $a_m(t)=a_m(0)\;e^{-i\mu t}$ in
the three Lagrange equations. A detailed analysis of the resulting $3\times 3$ nonlinear 
system shows that two classes of solution exist. The first class ($f_a$) corresponds
to $a_1=0$. Depending on the value of the parameters $Ng$, $A$ and $\Omega$, there
may exist two, three or four solutions of this kind. After some tedious but
straightforward calculation, one can obtain for this first class of solution an
analytical relation between $\Omega$ and the angular momentum per particle
$L=2|a_2|^2$:
 \begin{equation}
\Omega=1-\frac{Ng}{8\pi}\left(1-\frac{3}{8}L\right) \pm \sqrt{2}\,
A\frac{1-L}{\sqrt{L(2-L)}}\ .
\nonumber
 \label{eq:firstsolution}
 \end{equation}
The second class of solution corresponds to a non-zero value for $a_1$ and we have
not been able to provide an exact analytical expression for the solution in this
case. Using a numerical analysis, we have determined the local minima of the energy
and we found that one solution of this kind exists if and only if $\Omega>0.766$. We
have compared the energy of this solution with the lowest energy of the solutions in
the first class: for $\Omega<\tilde{\Omega}=0.773$ (respectively $\Omega>\tilde{\Omega}$),
the  the ground state is obtained with a solution belonging to the first (respectively second) class.

The stability of the solutions of the first class ($a_1=0$) can be studied
analytically by looking at the equation of evolution of $b_1=a_1e^{i\mu t}$. This
equation can be linearized around $b_1=0$ and written in the form $i\dot
b_1=Ab_1+Bb_1^*$, where the constants $A$ and $B$ are real numbers that can be
calculated explicitly in terms of the parameters $\Omega$, $A$ and $Ng$. The
stationary solution corresponds to $b_1=0$, and it is stable if $b_1(t)$ stays
around 0 when starting from a small non-zero initial value. This happens when
$|A|>|B|$, whereas $b_1$ undergoes an exponential divergence from any initial noise
if $|A|<|B|$, signalling a dynamical instability of the solution.

\noindent{\bf Acknowledgements}
We acknowledge discussions with I. Cirac and support of the EU SCALA and ESF Fermix Programs,
Spanish MEC grants (FIS 2005-03169/04627, QOIT) and the French
programs ANR and IFRAF.\\
\noindent{\bf Authors contributions}
All authors have contributed equally to this work.\\
\noindent{\bf Additional information}
Reprints and permissions information is available online at npg.nature.com/reprintsandpermissions.
Correspondence and requests for materials should be
addressed to N. B.


\begin{thebibliography}{99}


\bibitem{Weiss:1907}
Weiss, P. L'hypoth\`ese du champ mol\'eculaire et la propri\'et\'e ferromagn\'etique. {\it J. Phys.Th\'eor. et Appliq.} ${\bf 6,}$ 661-690 (1907).

\bibitem{pit}
Pitaevskii, L. \& Stringari, S. {\it Bose-Einstein Condensation}. (Oxford University Press, Oxford, 2003).

\bibitem{Jaksch:1998}
Jaksch, D., Bruder, C., Cirac, J.I., Gardiner, C.W. \& Zoller, P. Cold bosonic atoms in optical lattices. {\it Phys. Rev. Lett.} ${\bf 81,}$ 3108-3111 (1998).

\bibitem{coo}
Cooper, N.R. Rapidly rotating atomic gases. {\it Adv. Phys.} $\bf 57,$ 539-616 (2008).

\bibitem{Yoshioka:2002}
Yoshioka, D. {\it The Quantum Hall Effect} (Springer, 2002).

\bibitem{gri}
Griffin, A. {\it Excitations in a Bose-Condensed Liquid}, (Cambridge University Press, Cambridge, 1993).

\bibitem{fet}
Fetter, A.L. Rotating trapped Bose-Einstein condensates. {\it Laser. Phys.} $\bf 18,$ 1-11(2008).

\bibitem{Feder:2000}
Feder, D. L., Clark, C.W., \& Schneider, B.I. Nucleation of vortex arrays in rotating anisotropic Bose-Einstein condensates, {\it Phys. Rev. A} $\bf{61,}$ 011601(1-4) (2000).

\bibitem{sin}
Sinha, S., \& Castin, Y. Dynamic instability of a rotating Bose-Einstein condensate. {\it Phys. Rev. Lett.} ${\bf 87,}$ 190402 (2001).

\bibitem{kas}
Kasamatsu, K., Tsubota, M. \& Ueda, M. Nonlinear dynamics of vortex lattice formation in a rotating Bose-Einstein condensate. {\it Phys. Rev. A} ${\bf 67,}$ 033610 (2003).

\bibitem{but}
Butts, D.A., \& Roksar, D.S. Predicted signatures of rotating Bose-Einstein condensates. {\it Nature} $\bf{397,}$ 327-329 (1999).

\bibitem{Bertsch:1999}
Bertsch, G.F. \& Papenbrock, T. Yrast line for weakly interacting trapped bosons, {\it Phys. Rev. Lett.} ${\bf 83,}$ 5412-5414 (1999).

\bibitem{Smith:2000}
Smith, R.A., \& Wilkin, N.K. Exact eigenstates for repulsive bosons in two dimensions. {\it Phys. Rev. A} ${\bf 62,}$ 061602(1-4) (2000).

\bibitem{Jackson:2000}
Jackson, A.D. \& Kavoulakis,G.M. Analytical results for the interaction energy of a trapped, weakly interacting Bose-Einstien condensate. {\it Phys. Rev. Letters} ${\bf 85,}$ 2854-2856 (2000).

\bibitem{dag}
Dagnino, D., Barber\'an, N., Osterloh, K., Riera, A. \& Lewenstein, M. Symmetry breaking in small rotating clouds of trapped ultracold Bose atoms. {\it Phys. Rev. A} ${\bf 76,}$ 013625 (2007).

\bibitem{roma}
Romanovsky, I., Yannouleas, C., \& Landman, U. Symmetry-conserving vortex clusters in small rotating clouds of ultracold bosons. {\it Phys. Rev. A} $\bf{78,}$ 011606(R) (2008)


\bibitem{parke}
Parke, M.I., Wilkin, N.K., Gunn, J.M.F. \& Bourne, A. Exact vortex nucleation and cooperative tunneling in dilute BECs. {\it Phys. Rev. Lett.} $\bf{101,}$ 110401 (2008).

\bibitem{pit1}
Pitaevskii, L.P. Vortex lines in an imperfect Bose gas. {\it Sov. Phys. JETP} ${\bf 13,}$ 451-454 (1961).

\bibitem{gro}
Gross, E.P. Structure of a quantized vortex in boson systems {\it Nuovo Cimento} ${\bf 20,}$ 454-477
(1961).

\bibitem{strin99}
Stringari, S. Phase Diagram of Quantized Vortices in a Trapped Bose-Einstein Condensed Gas. {\it Phys. Rev. Lett.} ${\bf 82,}$ 4371-4375  (1999).

\bibitem{Ued}
Ueda, M. \& Nakalima, T. Nambu-Goldstone mode in a rotating Bose-Einstein condensate. {\it Phys. Rev. A} ${\bf 73,}$ 043603 (2006).

\bibitem{mor}
Morris, A.G. \& Feder, D.L. Validity of the lowest-Landau-level approximation for rotating Bose gases. {\it Phys. Rev. A} ${\bf 60,}$ 033605 (2006).

\bibitem{Wilkin:2000}
Wilkin, N.K. \& Gunn, J.M. Condensation of "Composite Bosons" in a Rotating BEC. {\it Phys. Rev. Lett.} ${\bf 84,}$ 6-9 (2000).

\bibitem{eckert}
Eckert, K., Schliemann, J., Bru\ss, D., \& Lewenstein, M. Quantum correlations in systems of indistinguishable particles. {\it Ann. Phys. (N.Y.)} ${\bf 299,}$ 88 (2002).

\bibitem{zan}
Zanardi, P. Quantum entanglement in fermionic lattices, {\it Phys. Rev. A} ${\bf 65,}$ 042101 (2001).

\bibitem{nun}
Nunnenkamp, A., Rey, A.M., \& Burnett, K. Cat state production with ultracold bosons in rotating ring superlattices. {\it Phys. Rev. A} ${\bf 84,}$ 023622 (2008).

\bibitem{Messiah}
Messiah, A. {\it Quantum Mechanics}, chapter XVII, \S 13 (Courier Dover Publications, 1999).

\bibitem{perez}
Perez-Garc\'ia, V.M., Michinel, H., Cirac, J.I., Lewenstein, M. \& Zoller, P. Low energy excitations of a Bose-Einstein condensate: A time-dependent variational analysis. {\it Phys. Rev. Lett.} ${\bf 77,}$ 5320-5323 (1996).

\bibitem{garay}
Garay, L.J., Anglin, J.R., Cirac, J.I. \& Zoller, P. Sonic analog of gravitational black holes in Bose-Einstein condensates. {\it Phys. Rev. Lett.} ${\bf 85,}$ 4643-4647 (2000).

\bibitem{Fetter:2007}
Fetter, A.L. Lowest-Landau-level description of a Bose-Einstein condensate in a rapidly rotating anisotropic trap. {\it Phys. Rev. A} ${\bf 75,}$ 013620 (2007).


\end{thebibliography}
\end{document}